\documentclass[twoside,fleqn]{ActaStyle}
\usepackage{times,cite}

\usepackage{lscape}             
\usepackage{graphicx,epsfig}    
\usepackage[tbtags]{amsmath}    

\def\authorheading{C. Hanhart}

\def\shorttitle{Hadronic production of $\eta$--mesons ...}

\begin{document}
\hfill{\small FZJ--IKP(TH)--2005--33}

\pagerange{1}{12}

\title{%
Hadronic production of $\eta$--mesons: recent results and open questions}

\author{
C. Hanhart\email{c.hanhart@fz-juelich.de}
}
{
Institut f\"{u}r Kernphysik, Forschungszentrum J\"{u}lich GmbH,\\ 
 D--52425 J\"{u}lich, Germany }

\day{November 7th, 2005}

\abstract{%
I review recent insights and open
questions connected with the production of $\eta$--mesons from hadrons.
}

\pacs{%
}

\section{Introductory remarks}

In recent years a large number of high accuracy data was published on the
production of $\eta$--mesons in various reactions, such as $\pi N\to \eta N$
\cite{pin2etan}, $\gamma N\to \eta N$ \cite{gammaN}, $NN\to NN\eta$
\cite{nneta,calen1}, $pn\to d\eta$ \cite{calen}, $\gamma d\to NN\eta$
\cite{gammad}, $\gamma {^3}$He$\to \eta {^3}$He \cite{pfeiffer}, $pd\to \eta{^3}$He
 \cite{pdeta} as well as data on heavier nuclei that I will not discuss
here.

Unfortunately, at present theory is far from being equally accurate and thus
we are now pushed to identify the class of questions that can be addressed
theoretically at present in a controlled way and to investigate those.
In this article I will present my personal point of view of what should be
studied in the coming years in respect to $\eta$ production from hadrons. The
eta--network provides the ideal soil for this enterprise.

It is well known that the production of eta mesons from single nucleons is
dominated by the resonance $S_{11}(1535)$ irrespective of the probe. Thus
investigating $\eta$ production off hadrons means to some extend to study the
$S_{11}$ in various settings. Especially since the nature of this resonance is
heavily debated in the literature (see also next section and references given
there), a systematic study of $\eta$ production in various environments is of
high importance.

\section{Production from single nucleons and the $S_{11}$}

Unfortunately it is not yet possible to derive hadron spectra from QCD
directly, although large progress has been made in lattice approaches recently
(see, e.g., Refs. \cite{lee}). We therefore do not yet understand how nature produces
hadrons out of quarks. The probably most prominent example that highlights our
degree of ignorance is the spectrum of the lightest baryons.

\begin{figure}[t]
\begin{center}
  \includegraphics[angle=0,width=.35\textwidth,height=!]{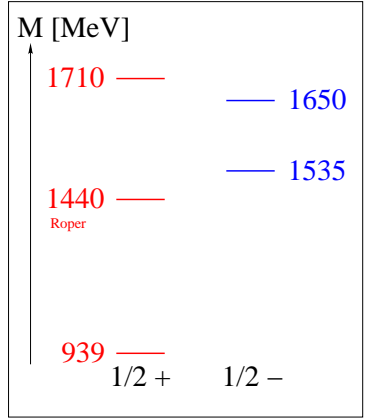}
\caption{Lightest baryons for $J^P=\frac12^+$ and  $J^P=\frac12^-$.}
\end{center}
\label{basp}
\end{figure}

The problem is reflected in the ordering of the various
  states: The non--relativistic quark model in its original formulation --- based
  on a harmonic oscillator potential \cite{isgur} --- predicts an arrangement of
  states with alternating parities. However, in the baryon spectrum the first
  positive parity excitation of the nucleon (the so called Roper resonance) is
  lighter than the first negative parity excitation --- the $S_{11}(1535)$. This
  is illustrated in Fig. \ref{basp}

The inclusion of an instanton induced interaction in the quark--quark
potential improved the picture, however, without changing the order of the
lightest states \cite{bernard}.  To my understanding there are so far two
possible explanations of the spectrum in the literature, one, where the Roper
is generated dynamically and thus interpreting the $P_{11}(1710)$ as the first
quark state \cite{schuetz,krehl}, and another, where the ordering of states was
changed by introducing a flavor dependent quark--quark interaction \cite{glozmanriska}.

At present it is not clear what the connection between these two pictures
is --- if such a connection exits at all. Recently the whole situation became
even more confusing for it was claimed that based on chiral dynamics the
$S_{11}(1535)$ appears as a dynamically generated state
\cite{kaiserweise,oset}.  A  look at the baryon spectrum --- Fig. \ref{basp} --- reveals
that, if confirmed, this puts the quark model in a complicated situation for
now the order of states is in even more severe disagreement with its
predictions and it is hard to imagine that a flavor dependent interaction can
overcome the huge gap between the 1650 and the 1440. It is therefore of high
importance to understand the nature of the low lying resonances for this
promises deep insights into how the nature makes hadrons.

How is it possible to distinguish molecular states and compact quark states? Based on an
old proposal by Weinberg \cite{wein} in Ref. \cite{f0nat}, in line with a
series of older works \cite{natold}, we identified the
energy dependence of the elastic scattering amplitude in the molecule forming
channel as a key quantity. The method applies if the binding energy of the
state is significantly smaller than any other scale of the problem. In
particular the inelastic threshold needs to be far away. Thus, to understand
if the $S_{11}$ is an $\eta N$ molecule, experimental information is needed on
elastic $\eta N$ scattering. Unfortunately this channel is not directly
accessible. Various theoretical analyses of the non--diagonal channels lead to
only very weak constraints of, e.g., the $\eta N$ scattering length: a
compilation presented in Ref. \cite{sonja} gives
\begin{equation}
a_{\eta N}=(0.2-1.1,0.26-0.35) \ \mbox{fm} ,
\label{etanscat}
\end{equation}
where the first range refers to the real part and the second to the imaginary part. 
An imaginary part to a scattering length arises in the presence of inelastic
channels.
Here these are the $\pi N$ and the $\pi\pi N$ channels.
 An improved analysis
using all available data is therefore urgently called for. Note that an
updated analysis for $\pi N$ scattering is currently under way \cite{peko}.

\begin{figure}[t!]
\begin{center}
  \includegraphics[angle=0,width=.6\textwidth,height=!]{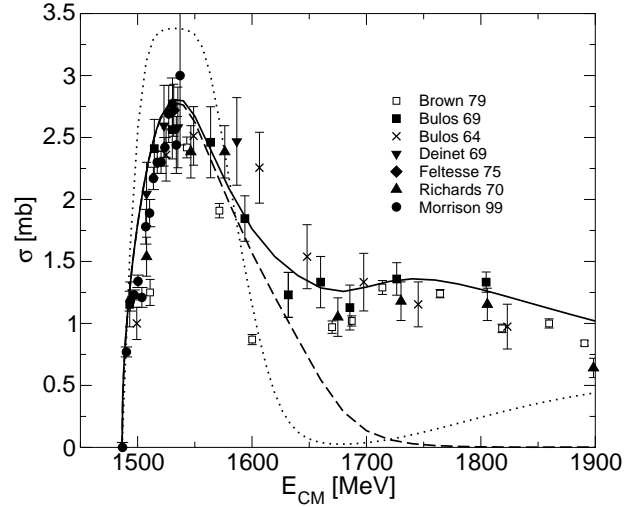}
\caption{Comparison of the measured cross section for $\pi N\to \eta N$ to
  various calculations. The dotted line denotes the results of Ref.
  \cite{krehl}, where no direct coupling of the $S_{11}$ to the two pion
  channels was included. The dashed line shows the results of Ref.
  \cite{gasparyan} --- where a coupling of the $S_{11}$ to the $\pi \Delta$
  channel in a $d$--wave was included --- for this cross section for only the
  $S_{11}$ channel (dashed line) and for the full result (solid line).
The data was taken from Refs. \cite{pin2etan}.}
\label{piN2etaN}
\end{center}
\end{figure}

 Experimental information on the $S_{11}$ can be derived from
data on $\pi N\to \pi N$, $\pi N\to \eta N$ and $\pi N\to \pi\pi N$ as well as
the corresponding $\gamma$ induced reactions to control systematics. One may
ask whether the two pion channels are really necessary --- after all the
particle data booklet only lists a branching ratio of 1--10 \% for all the two
pion channels. However, a direct comparison of the total cross section for
$\pi N\to \eta N$ and the inelasticity of $\pi N$ scattering in the $S_{11}$
channel reveals a need for these. Using the optical theorem,
one can directly convert the inelasticity $\eta$ to the inelastic cross
section. Based on the values for
momentum and inelasticity as given in Ref. \cite{arndt} we find at  $E_{tot} = 1540$ MeV 
$\sigma_{in}^{\pi N}{=}{2\pi}(1-\eta^2)/3k_1^2{=}3.5 \ \mbox{mb}$,
where $k_1$ denotes the incoming pion momentum.  If the $\eta N$ channel would
be the only inelastic channel that couples to $\pi N$ in the $S_{11}$--partial
wave, these 3.5 mb need to agree to the peak value of the $\pi N \to \eta N$
cross section. However, the measured value is significantly lower. The
situation is illustrated in Fig. \ref{piN2etaN}. Thus a proper inclusion of
the two pion channels is necessary, as was stressed in
Refs. \cite{ioune,gasparyan,baru}.

To summarize the investigations based on reactions on a single baryon we state
that in order to better understand the nature of the $S_{11}(1535)$ both high
accuracy data and theoretical studies are necessary to derive constraints on
the $\eta N$ scattering parameters.

In addition, it is expected that a (loosely bound) molecule and a compact
quark state behave very different in the presence of additional baryons. This
field is unfortunately still lacking a systematic approach at present,
although microscopic calculations are available for both the $\eta
d$~\cite{pena} and the $\eta {^3}$He~\cite{etanuctheo} system. In what follows
I will briefly present and discuss on a qualitative level some phenomena
recently observed in theses systems.

\begin{figure}
\begin{center}
    \includegraphics[angle=0,width=.6\textwidth,height=!]{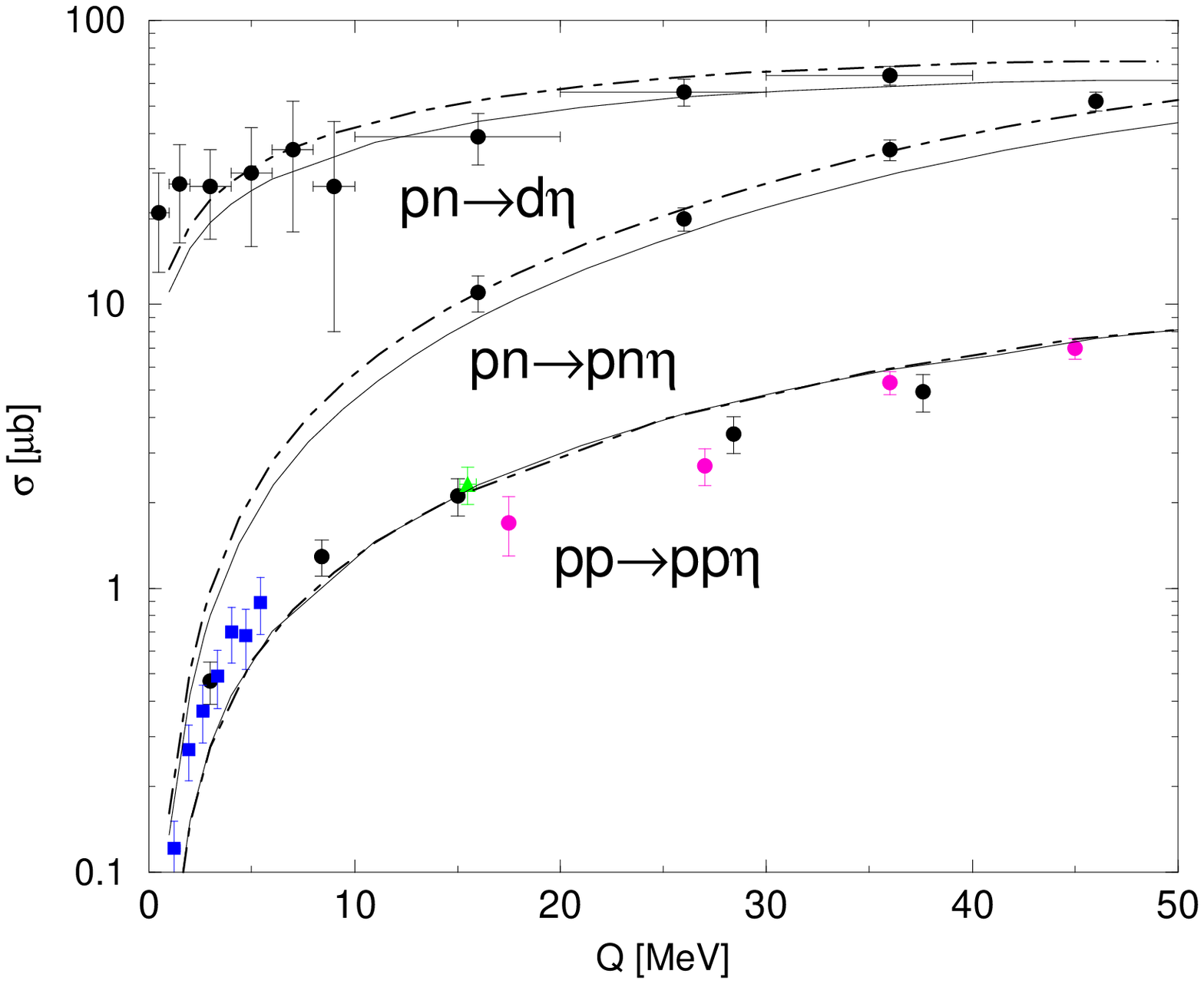}
\caption{Energy dependence of the total
    cross sections for the  
various 
$\eta$ production channels. The different curves are results employing
    different $NN$ wave functions in a microscopic calculation for the
    different channels. The picture is taken from Ref. \cite{baru}.}
\label{endeptotal}
\end{center}
\end{figure}

\section{The reaction $NN\to \eta NN$}

Unpolarized Data is available for total and differential cross sections for
$pp\to pp\eta$ \cite{nneta} and for total cross sections for $pp\to pn\eta$
\cite{calen1} and $pp\to d\eta$ \cite{calen}. For $pp\to pp\eta$
analyzing powers were measured as well \cite{winter}.

The energy dependence of the total cross sections for the various eta
production channels in $NN$ collisions is shown in Fig. \ref{endeptotal}.  The
curves are the results of a microscopic calculation using two different models
for the $NN$ interaction \cite{baru}. This calculation includes the $\eta
N$ interaction only to leading order. The picture nicely illustrates that it
is necessary to go beyond perturbation theory for the $\eta NN$ system to
understand the energy dependence at low energies. This was done in Ref.
\cite{arenh} for the channel $pp\to pp\eta$ and in Refs. \cite{veras,pena} for
$pp\to d\eta$ and it was indeed possible to describe the energy dependence of
the various channels, once few body equations were employed for the final
state interaction.

\begin{figure}[t!]
\begin{center}
  \includegraphics[angle=0,width=.6\textwidth,height=!]{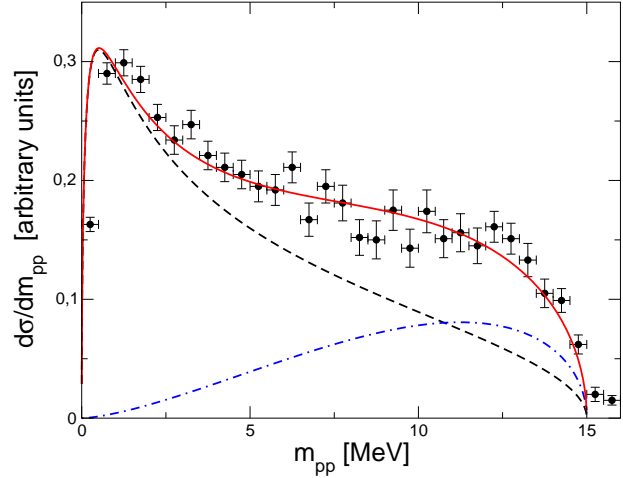}
\caption{Plot of the $pp$ invariant mass spectrum for the reaction $pp\to
  pp\eta$ at an excess energy of 15 MeV. The dashed line shows the
  distribution expected, if only the $pp$ $S$--wave contributes (including the
  $NN$ FSI). As an example, the dot--dashed curve shows the expected distribution for $NN$
  $P$--waves and the solid line is the sum of both. The data is from Refs.
  \cite{nneta}.}
\label{etaspec}
\end{center}
\end{figure}

However, it turned out that the $pp$ invariant mass spectrum measured for
$pp\to pp\eta$ \cite{nneta} can not be described in this way --- the data is
shown in Fig. \ref{etaspec}. The calculation that was capable to describe the
energy dependence of the total cross section quite well, failed to describe
the invariant mass spectrum \cite{arenh}: The model results for this
observable were quite close to the dashed line in the figure that was
calculated under the assumption that only the $NN$ FSI distorts the $NN$
$S$--waves. To resolve the discrepancy it was speculated that either there is
a significant $\eta N$ interaction \cite{pawel}, not captured by the model of
Ref. \cite{arenh}. On the contrary, Deloff argued that there might be a
significant energy dependence of the production operator \cite{delhoff}, and
in Ref.  \cite{kanzo} a significant contribution of $NN$ $P$--waves were
proposed as a possible solution for the puzzle. Note that even isotropic
angular distributions can be compatible with $NN$ $P$--waves.  All the can be
read of the $pp\eta$ invariant mass spectrum is that a term quadratic in the
final momentum is missing to explain the data (c.f. Fig. \ref{etaspec}).
However, such a dependence would emerge from all three explanations.
Fortunately it is possible to isolate the $NN$ $P$--waves
by a double polarized measurement. For this it is sufficient to measure
$A_{xx}$ as well as the differential cross section $\sigma_0$, since
$$
^3\sigma_\Sigma = \sigma (1+A_{xx})=2\sigma (\uparrow \uparrow) \ ,
$$
where the arrows indicate that both the spin of the beam as well as the target
are aligned perpendicular to the beam.
Consequently the spin singlet initial state --- and therefore neither $^1S_0\to
^3P_0s$ nor $^1D_2\to ^3P_2s$ --- does not contribute to
$^3\sigma_\Sigma$\footnote{Here we use the notation $^{2S+1}L_J\to
  ^{2S'+1}L'_{J'} l_\eta$, where $S$, $L$, $J$ ($S'$, $L'$, $J'$) denote 
  spin, angular momentum, and total angular momentum of the initial (final)
  $NN$ pair; $l_\eta$ denotes the angular momentum of the outgoing
  $\eta$--meson with respect to the $NN$ system. For a review of the selection
rules see Ref. \cite{report}.}. Note
the observable is the same that can be used to
  measure the parity of narrow resonances \cite{thetapar}.

It should be clear that only when the partial wave decomposition of the
$pp\eta$ spectrum is known, we can understand the role the $\eta N$
interaction plays here. Especially that the calculation of Ref. \cite{arenh}
failed to describe the spectrum indicates that a lot is still to be learned
from the reactions $NN\to NN\pi$. For example, in Ref. \cite{kanzo} it is
shown that the requirement to describe the $pp\eta$ spectrum by $NN$
$P$--waves strongly constraints the $NN\to NN^*(1535)$ transition potential.

\section{The reactions $pd\to \eta {^3}$He and $\gamma{^3}$He$\to \eta{^3}$He}

The most prominent effect of the $\eta$--few--nucleon interaction can be seen
in $\eta$--nucleus systems, like $\eta d$ \cite{calen}, $\eta ^3$He
\cite{pdeta}, and $\eta \alpha$ \cite{ola}.

Phenomenological investigations \cite{haider,phenoetanuc,green} as well as
microscopic calculations for $\eta$--nucleus interactions exist from various
groups \cite{etanuctheo} --- some of these will be presented  in the
contribution by T. Pena to these proceedings. Thus here I will more focus on
those features of the systems $\eta d$ and $\eta ^3$He that can be directly
read off the data. The tool we use is a final state interaction enhancement
factor $f(q)$, where $q$ denotes the momentum of the final state particles
in the center of mass,
according to Watson and Midgal \cite{watson}
 in the scattering length approximation
\begin{equation}
f(q) = \frac1{1-iaq}=\frac1{1-ia_Rq+a_Iq} \ ,
\label{fsifac}
\end{equation}
where $a_R$ ($a_I$) denotes the real (imaginary) part of the $\eta$--nucleus scattering
length.  As in case of the elementary amplitude (c.f. Eq.
(\ref{etanscat})) $a$ is a complex valued quantity because of the presence
of inelastic channels.  Note that one should be careful in
interpreting the value of $a$ extracted from a fit to production data as the
scattering length. It is known since long, that there is a well defined
connection between the parameters of elastic scattering and the effect of
final state interactions.  This connection is put on a solid theoretical basis
using dispersion theory \cite{goldberger}. However, a direct proportionality
between the energy dependence of scattering and production only holds, if the
scattering length is significantly larger than the next term in the effective
range expansion --- the effective range.  In addition, a systematic study of
final state interaction effects revealed that the Watson formula tends to give
a scattering length that is too large; however, there is a clear
correlation amongst the values of elastic scattering and production
\cite{achot2}.  One should also stress that in the presence of inelasticities
it is not even possible to derive an expression for the final state
interaction effect in closed form. Thus the results should be taken on a
qualitative level.

\begin{figure}[t!]
\begin{center}
 {\includegraphics[angle=0,width=.48\textwidth,height=8cm]{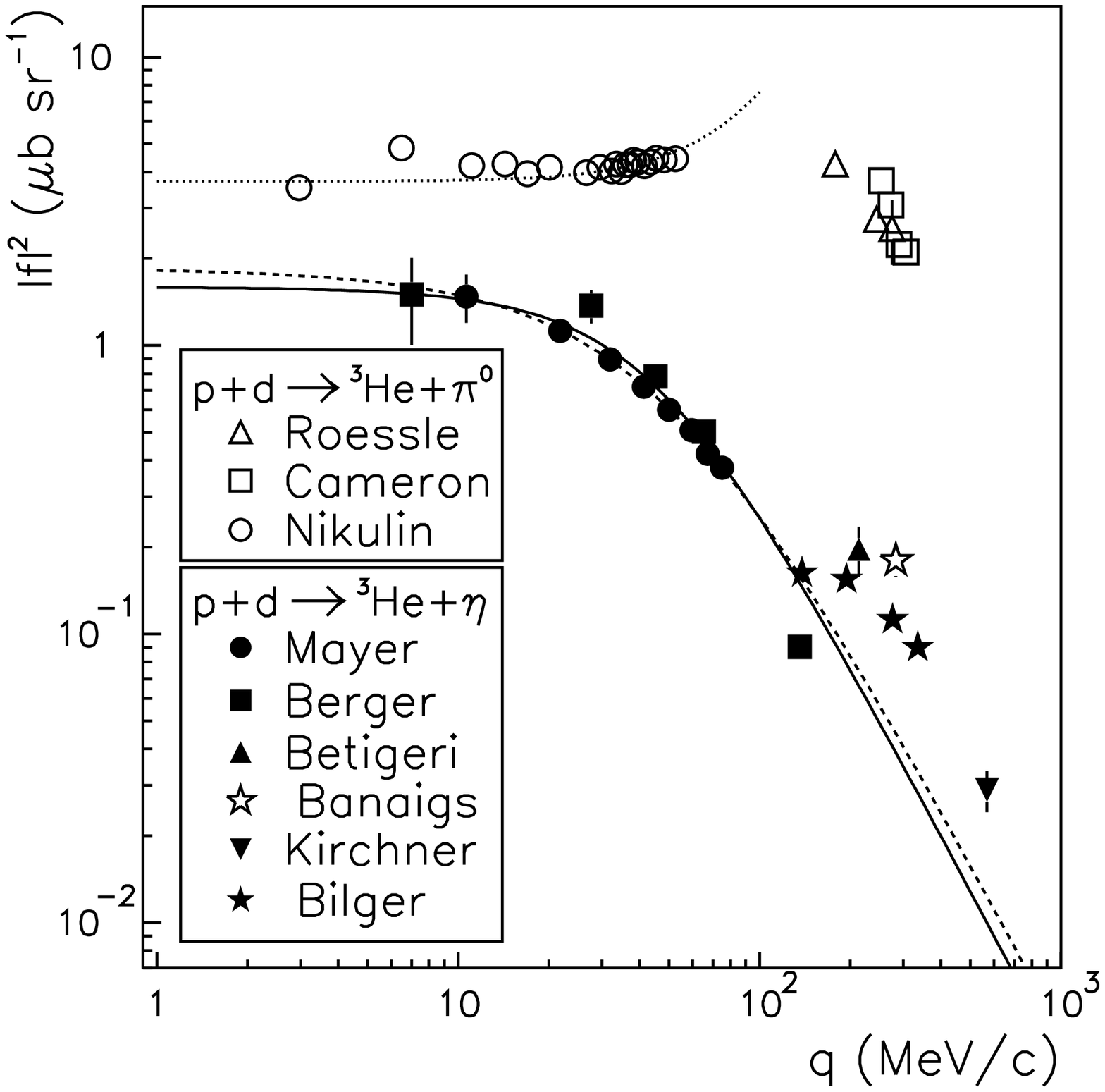}}
{\includegraphics[angle=0,width=.48\textwidth,height=8cm]{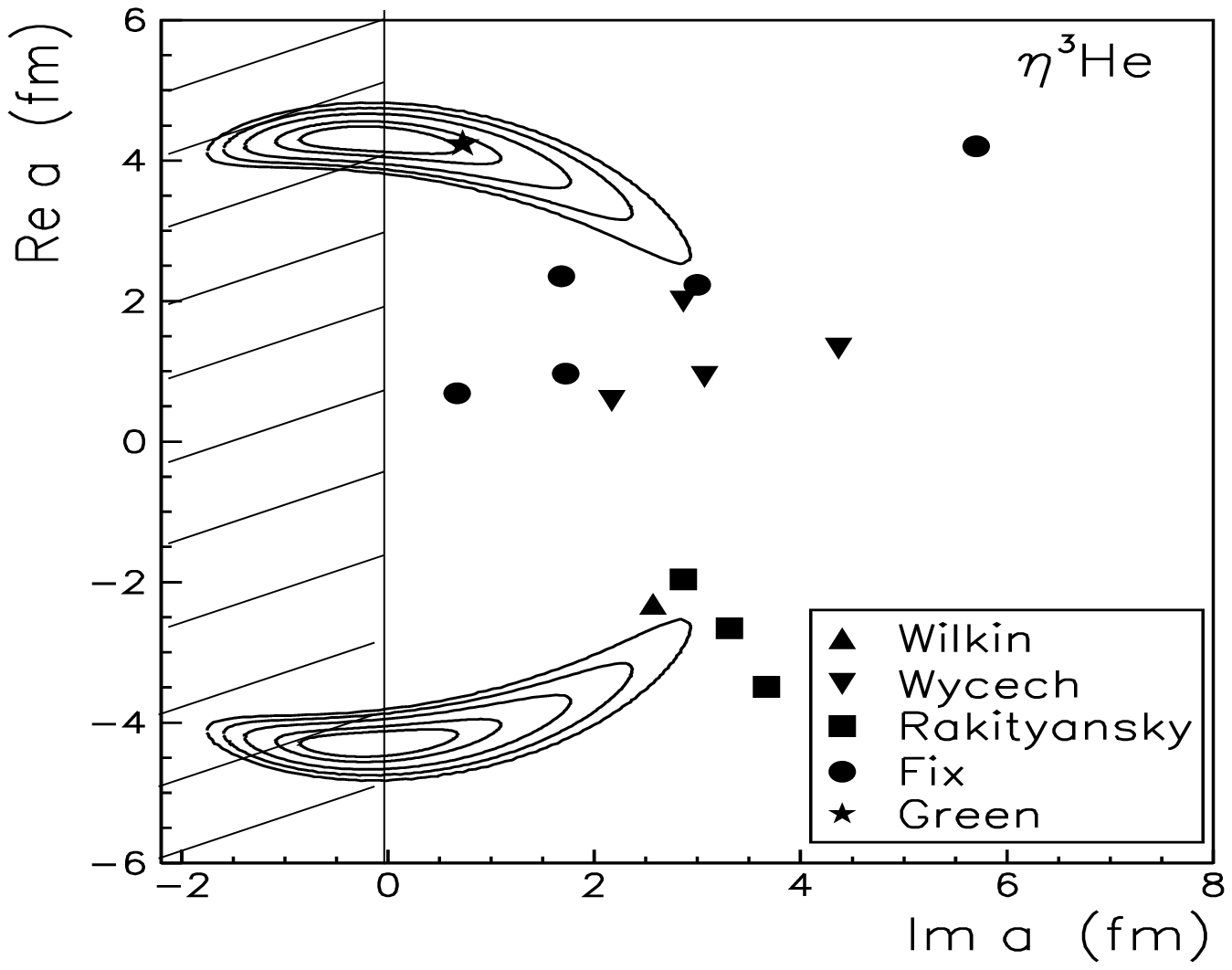}}
\caption{Result for the fit of the FSI formula of Eq. (\ref{fsifac}) to the
  world data on the reaction $pd\to \eta {^3}$He. The figures are taken from
  Ref. \cite{saschas}, where also a complete list of experimental references
  is given. The left panel the data for the amplitude is shown as well as our
  best fits. For comparison also data on $pd\to \pi^0 {^3}$He is given. The
  left panel shows the confidence levels for the extracted real and imaginary
  part of the scattering length.}
\label{etahefit}
\end{center}
\end{figure}

To study the singularity structure of Eq. (\ref{fsifac}) we need to
analytically continue the momentum into the complex plane. The physical sheet
is given by those momenta with positive imaginary parts --- the unphysical one
by negative imaginary parts.  The sign of the imaginary part of the scattering
length, on the other hand, is fixed to be positive through unitarity. It is
the sign of the real part that decides, whether the nearby pole we want to
investigate refers to a bound state or a virtual state. Thus only a negative
real part refers to a bound state\footnote{Unfortunately there are different
  sign conventions present in the literature. We here use that of Goldberger
  and Watson \cite{goldberger}, which is common for meson--nucleon
  systems. Note, however, for nucleon--nucleon scattering the scattering
  length is traditionally defined with a relative minus sign compared to the
  convention used here.}. Haider and Liu \cite{haider} pointed out that for
the existence of a bound state the additional condition $a_I<|a_R|$ is to hold.

How can we measure the sign of the imaginary part? In a cross section
measurement what typically enters observables close to the threshold is
$|f(q)|^2$. Then, above threshold, we get
$$|f(q)|^2=\frac{1}
{1+2a_Iq+|a|^2q^2} \ , \quad \mbox{with} \ q=\sqrt{2\mu E} \, ,$$
where $E$ denotes the kinetic energy of the final system with respect to the
$\eta$--nucleus threshold and $\mu$ is the corresponding reduced mass.
Therefore above threshold $E>0$. 
Thus any measurement above the $\eta$--nucleus threshold is sensitive only to
the magnitude of the real part, but not to its sign;
measurements above threshold are necessary  to pin down the absolute
values of the real  and the imaginary part of the scattering length. It
should be clear from the given formula that only measurements very close to
the threshold allow one to disentangle $a_I$ from $a_R$.

On the other hand, below threshold  $(E\le 0)$ we get
$$
|f(q)|^2=\frac{1}{1+2a_R\kappa+|a|^2\kappa^2} \ , \quad \mbox{with} \ 
\kappa=\sqrt{-2\mu E}\, .$$
Thus, now real and imaginary part have changed
their roles. Since we know the sign of the imaginary part high accuracy
measurements of the energy dependence of the $\eta$--nucleus amplitude above
and below the eta threshold allows one to extract the sign of the real part of
the scattering length \cite{mixingpaper}.

But what does it mean to measure an amplitude below threshold? This can only
work by identifying inelastic channels that show a significant coupling to the
$\eta$--nucleus channel. We will discuss one example for this in great detail
below. However, we would first like to briefly comment on the expected
magnitude of the imaginary part. We know that there is a significant
coupling of $\pi N \to \eta N$ (c.f. Eq. (\ref{etanscat})) near threshold. One
might therefore expect an imaginary part of the scattering length of the order of at
least one
fermi. In contrast to this in a recent analysis of the world data base for $\eta
{^3}$He values for the imaginary part of the $\eta {^3}$He scattering length
were extracted that were compatible with zero \cite{saschas}. The result of
this fit is illustrated in Fig. \ref{etahefit}. The left panel shows the
scattering amplitude, defined through
\begin{equation}
{|f(q)|^2=(q/k)\sigma_{tot}/4\pi} \ ,
\end{equation}
where $k$ denote the initial cms momentum.
From this fit we extracted
\begin{equation}
a{=}|4.3{\pm}0.3|{+}i(0.25{\pm}0.25) \ \mbox{fm} \ .
\label{etahe}
\end{equation}
The numbers given are in line with a recent K--matrix analysis \cite{green}.

Can we understand such a small imaginary part of the scattering length? 
The answer is yes. Let us start the discussion for simplicity with the $\eta
d$ system.
The central observation was that the Pauli Principal for
 few--nucleon systems also needs to hold for intermediate states
 \cite{recoils}. Technically this is to be realized by a consistent inclusion
 of self energy diagrams and rescatterings. To understand the role of the  $\pi NN$
 intermediate state in the regime of a prominent $\eta d$ $s$--wave
 interaction, we observe, that the $\eta N\to \pi N$ operator must be an
 isovector acting on the nucleons. On the other hand, a $\eta N$ $s$--wave
 necessarily connects to a $\pi N$ $s$--wave and therefore the operator is
 spin independent. If such an operator acts on the deuteron wave function it
 forces the $NN$ pair into an isospin 1 state, but leaves the spin in the spin
 triplet. A spin triplet $NN$ pair with isospin 1, however, has necessarily
 odd angular momentum and therefore, to conserve parity, also the pion must be
 in a $p$--wave. As a consequence the potentially most prominent inelastic
 channel is blocked. As the deuteron is a prominent building block also of
 $^3$He this argument at least to some extend should hold as well.
This was worked out in more detail in Ref.  
\cite{jounis}.

\begin{figure}
\begin{center}
    \includegraphics[angle=0,width=.6\textwidth,height=!]{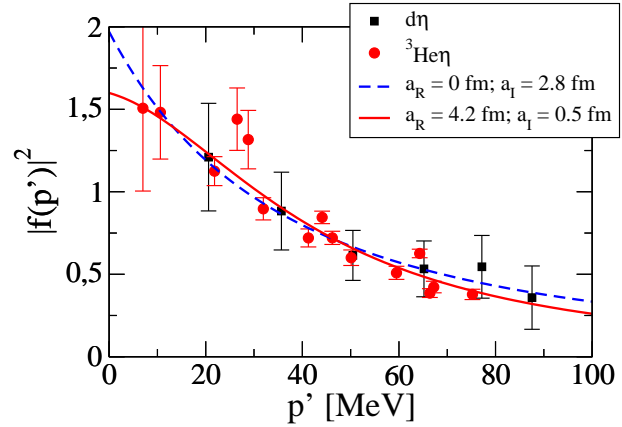}
\caption{Energy dependence of the $\eta d$
      (squares) and $\eta ^3$He (dots) interaction as derived from
      cross section data of the reactions $pn\to d\eta$ \cite{calen} and
      $pd\to\eta ^3$He \cite{pdeta}. The dashed line corresponds the best
      fit of the final state interaction formula Eq. (\protect\ref{fsifac})
      solely to the $\eta d$ data, whereas the solid curve corresponds to the
      best fit to the $\eta ^3$He data.}
\label{etadetahe}
\end{center}
\end{figure}

In the process of preparing this talk I observed an amusing similarity that I
wanted to share. When plotted in the same figure both the $\eta d$ interaction
and the $\eta {^3}$He interaction show a very similar energy dependence. This
is illustrated in Fig. \ref{etadetahe}. The curves show a best fit to the
$\eta d$ data as well as the best fit to the $\eta {^3}$He data, as explained above.
This similarity might point at the conjecture used above, namely that the
$\eta {^3}$He dynamics is at least to a large extend given by the interaction
of the $\eta$ with the deuteron substructure of the He.

In contrast to the imaginary part of the scattering length, the real part as extract
from the fit to the $pd\to \eta {^3}$He data is quite sizable. Under the
assumption that this large value is a signal of a bound state, in
Ref. \cite{sascha2} the relation between the scattering length and the
position of the bound state pole was investigated and values quite close to
the threshold were extracted. However, to make this analysis useful we need to
know, if a bound state exists or not.

Therefore we
 now turn to $\eta$--nucleus measurements below the $\eta$ threshold. As
mentioned here we have to look at different channels. Experiments for this
were proposed quite some ago \cite{askalbrecht}, but only realized recently.
At COSY there was a measurement of the ratio $pd\to \pi^+t$ to $pd\to \pi^0
{^3}$He \cite{magiera} and it could be shown that this ratio is indeed
sensitive to the sign of the real part of the $\eta$--nucleus scattering
length \cite{mixingpaper}. Since the existing data is not of sufficient accuracy
and there is currently no new measurement planed, we will not describe this
here in detail. 

\begin{figure}
\begin{center}
    \includegraphics[angle=0,width=.6\textwidth,height=!]{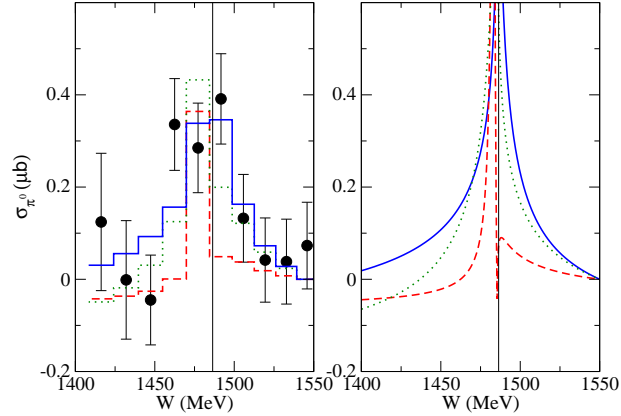}
\caption{Comparison of the various fit to the data, as a function of the
  reduced photon energy $W$ defined in Ref. \protect\cite{pfeiffer}. 
 The left panel:
  binned as the data; the right panel:
  no binning. The solid (dashed) line corresponds to $a=(+4,1)$ ($a=(-4,1)$)
~fm,
 and the dotted one to $a=(0,3.5)$~fm.  The vertical line 
 indicates the position of the $\eta {^3}$He threshold.}
\label{pfeifferfig}
\end{center}
\end{figure}

The TAPS collaboration at MAMI recently ran a quite successful experiment that
clearly showed a strong $\eta {^3}$He interaction right below threshold
\cite{pfeiffer}. What was measured was the reaction $\gamma {^3}$He$\to \pi^0
pX$, where, in order to find the signal, a cut was introduced to only count
$\pi^0 p$ pairs that go out back--to--back in the center of mass system. This
cut was motivated by the observation that if there were a bound
$\eta$--nucleus system the $S_{11}(1535)$ would certainly play a prominent
role in it. As the measurement was performed right at threshold this $S_{11}$
should be at rest and correspondingly the outgoing $\pi^0$--$p$ pairs from its
decay should go out back--to--back in the cms of the whole system. And indeed
a clear peak structure could be identified in the data. The structure
extracted, after subtraction of the neighboring angular bins, is shown in the
left panel of Fig. \ref{pfeifferfig} as a function of the reduced photon
energy. The perpendicular line indicates the position of the $\eta {^3}$He
threshold. To analyze this data we may use the same formula for the final
state interaction effects as used for other production reactions --- Eq.
(\ref{fsifac}) --- with both imaginary part and real part of the scattering
length as extracted from $pd\to \eta {^3}$He (c.f. Eq. (\ref{etahe})). This is
possible because final state interactions are universal in large momentum
transfer reactions.

There is one additional comment necessary before
we can  apply Eq. (\ref{fsifac}) to the TAPS data: there is in principle some
interference with the background possible. Thus, what was identified as the resonance
signal might well have some contribution from an interference term, and the
full signal may be written as
\begin{equation}
N\left(2\mbox{Re}(Bf^{res})+\left|f^{res}\right|^2\right) \ ,
\label{full}
\end{equation}
where $B$ is some complex number parameterizing that part of the background
that is allowed to interfere with the resonance signal and $N$ is a measure of
the total strength of the signal. Therefore, three different fits were
performed: fit 1 included only the pure resonance signal ($B=0$; only $N$ as a
free parameter); fit 2 included only the interference term ($B\to \infty$; $N$
and the phase of $B$ as a free parameter); and fit 3 considered the full
structure (thus here we have 3 free parameters: $N$, $|B|$ and the phase of
$B$). As it turned out, the $\chi ^2$ per degree of freedom for the two
scenarios (positive and negative real part of the scattering length) was
almost the same in all three cases and thus for illustration in Fig.
\ref{pfeifferfig} we only show the results of the second
fit, where the curves in the left panel correspond to the results after binning in
accordance with that of the experiment and the right panel corresponds to the
unbinned results.  To keep the number of free parameters low we choose
$a=(\pm 4,1)$ fm.  In both figures the dashed line corresponds to a negative
real part (indicating the existence of a bound state) and the solid line
corresponds to a positive real part (indicating a virtual state). 
For comparison also a curve is shown that has the maximum imaginary part
together with a vanishing real part still compatible with a subset of the
available $pd\to \eta {^3}$He data~\cite{saschas}.
The fit gave
a $\chi ^2$ per degree of freedom of 1 for the latter case, whereas it was
worse than 3 in the former.  Thus the data prefers the solution that
corresponds to a virtual state, although the existence of a bound state can
not be excluded, given the quality of the data.  Note, already in Ref.
\cite{sascha2} the interpretation of the TAPS data as a bound state was
questioned.
Fortunately a new experiment will be performed soon. The expected much higher
statistics promise for a near future an unambiguous decision on what scenario
is realized: a bound state or a virtual state.

\section{Summary}

In this talk various aspects of $\eta$--meson production were discussed.
The main issue is to investigate the $\eta$--nucleon interaction and in
particular the $S_{11}$ resonance in various environments. These studies
promise insights not only into the nature of this state but also into the
existence of meson--nucleus bound states.
Especially on the theory side a lot still needs to be done, but this field of
research promises deep insight in strong interaction physics in the
non--perturbative regime.

\vspace{0.5cm}

{\bf Acknowledgments}

Thanks to Pawel Moskal and his team of organizers for an educating, inspiring,
and entertaining workshop and to the network for making it possible --- and for
financial support. I also thank my collaborators on various aspects of the
work presented: V. Baru, J. Haidenbauer, A. Gasparyan, U.--G. Mei{\ss}ner, K.
Nakayama, J. Niskanen, A. Sibirtsev, and J. Speth.


\begin{thebibliography}{10}
\bibitem{pin2etan} R.~M.~Brown {\it et al.},
  Nucl.\ Phys.\ B {\bf 153} (1979) 89;
  F.~Bulos {\it et al.},
  Phys.\ Rev.\  {\bf 187} (1969) 1827;  W.~Deinet, H.~Mueller, D.~Schmitt, H.~M.~Staudenmaier, S.~Buniatov and E.~Zavattini,
  Nucl.\ Phys.\ B {\bf 11} (1969) 495;
 J.~Feltesse {\it et al.},
  Nucl.\ Phys.\ B {\bf 93} (1975) 242;
  W.~B.~Richards {\it et al.},
  Phys.\ Rev.\ D {\bf 1} (1970) 10.
 N.~C.~Debenham {\it et al.},
  Phys.\ Rev.\ D {\bf 12} (1975) 2545.
\bibitem{gammaN}
  F.~Renard {\it et al.}  [GRAAL Collaboration],
  Phys.\ Lett.\ B {\bf 528} (2002) 215 [arXiv:hep-ex/0011098]; B.~Delcourt,
  J.~Lefrancois, J.~P.~Perez-Y-Jorba, G.~Sauvage and G.~Mennessier,
  Phys.\ Lett.\ B {\bf 29} (1969) 75;
  B.~Krusche {\it et al.},
  Phys.\ Rev.\ Lett.\  {\bf 74} (1995) 3736.
\bibitem{nneta}
  P.~Moskal {\it et al.},
  Phys.\ Rev.\ C {\bf 69} (2004) 025203
  [arXiv:nucl-ex/0307005];
  M.~Abdel-Bary {\it et al.}  [COSY-TOF Collaboration],
  Eur.\ Phys.\ J.\ A {\bf 16} (2003) 127
  [arXiv:nucl-ex/0205016].
\bibitem{calen1}
  H.~Calen {\it et al.},
  Phys.\ Rev.\ C {\bf 58} (1998) 2667.
\bibitem{calen}
H.~Calen {\it et al.},
Phys.\ Rev.\ Lett.\ {\bf 79} (1997) 2642.
\bibitem{gammad}
   V.~Hejny {\it et al.},
  Eur.\ Phys.\ J.\ A {\bf 13} (2002) 493
  [arXiv:nucl-ex/0304005];
 B.~Krusche {\it et al.},
  Phys.\ Lett.\ B {\bf 358} (1995) 40;
 B.~Krusche {\it et al.},
  Phys.\ Lett.\ B {\bf 358} (1995) 40.
\bibitem{pfeiffer}
M.~Pfeiffer {\it et al.},
Phys.\ Rev.\ Lett.\  {\bf 92} (2004) 252001.
\bibitem{pdeta}
R.~Bilger {\it et al.},
  Phys.\ Rev.\ C {\bf 65} (2002) 044608.
   B.~Mayer {\it et al.},
  Phys.\ Rev.\ C {\bf 53} (1996) 2068;
M. Betigeri {\it et al.},
  Phys.\ Lett.\ B {\bf 472} (2000) 75.
\bibitem{lee}
 N.~Mathur {\it et al.},
  Phys.\ Lett.\ B {\bf 605} (2005) 137
  [arXiv:hep-ph/0306199];
  O.~Jahn, J.~W.~Negele and D.~Sigaev,
  arXiv:hep-lat/0509102.
\bibitem{isgur}
N. Isgur and G. Karl,
  Phys.\ Rev.\ D {\bf 18} (1978) 4187;
  Phys.\ Rev.\ D {\bf 19} (1979) 2653.
\bibitem{bernard}
 U.~L\"oring, B.~C.~Metsch and H.~R.~Petry,
  Eur.\ Phys.\ J.\ A {\bf 10}, 395 (2001)
  [arXiv:hep-ph/0103289].
\bibitem{schuetz}
  C.~Sch\"utz, J.~Haidenbauer, J.~Speth and J.~W.~Durso,
  Phys.\ Rev.\ C {\bf 57} (1998) 1464.
\bibitem{krehl}
  O.~Krehl, C.~Hanhart, S.~Krewald and J.~Speth,
  Phys.\ Rev.\ C {\bf 62}, 025207 (2000)
  [arXiv:nucl-th/9911080].
\bibitem{glozmanriska}
  L.~Y.~Glozman and D.~O.~Riska,
  Phys.\ Rept.\  {\bf 268} (1996) 263
  [arXiv:hep-ph/9505422].
\bibitem{kaiserweise}
  N.~Kaiser, P.~B.~Siegel and W.~Weise,
  Phys.\ Lett.\ B {\bf 362} (1995) 23
  [arXiv:nucl-th/9507036].
\bibitem{oset}
  T.~Inoue, E.~Oset and M.~J.~Vicente Vacas,
  Phys.\ Rev.\ C {\bf 65} (2002) 035204
  [arXiv:hep-ph/0110333].
\bibitem{wein}
 S.~Weinberg,
  Phys.\ Rev.\  {\bf 130} (1963) 776.
\bibitem{f0nat}
  V.~Baru, J.~Haidenbauer, C.~Hanhart, Y.~Kalashnikova and A.~Kudryavtsev,
  Phys.\ Lett.\ B {\bf 586} (2004) 53
  [arXiv:hep-ph/0308129].
\bibitem{natold} D.~Morgan,
  Nucl.\ Phys.\ A {\bf 543} (1992) 632;
Nils A. Tornqvist,  Phys. Rev. D {\bf 51} (1995) 5312.
\bibitem{sonja}
  A.~Sibirtsev  {\it et al.},
  Phys.\ Rev.\ C {\bf 65} (2002) 044007
  [arXiv:nucl-th/0111086].
\bibitem{peko} P.~Piirola and M.~E.~Sainio,
  Int.\ J.\ Mod.\ Phys.\ A {\bf 20} (2005) 1810.
\bibitem{arndt}  R.~A.~Arndt, I.~I.~Strakovsky, R.~L.~Workman and M.~M.~Pavan,
  Phys.\ Rev.\ C {\bf 52} (1995) 2120
  [arXiv:nucl-th/9505040].
\bibitem{ioune}  T.~Inoue, E.~Oset and M.~J.~Vicente Vacas,
  Phys.\ Rev.\ C {\bf 65} (2002) 035204
  [arXiv:hep-ph/0110333].
\bibitem{gasparyan}  A.~M.~Gasparyan, J.~Haidenbauer, C.~Hanhart and J.~Speth,
  Phys.\ Rev.\ C {\bf 68} (2003) 045207
  [arXiv:nucl-th/0307072].
\bibitem{baru}  V.~Baru, A.~M.~Gasparyan, J.~Haidenbauer, C.~Hanhart, A.~E.~Kudryavtsev and J.~Speth,
  Phys.\ Rev.\ C {\bf 67} (2003) 024002
  [arXiv:nucl-th/0212014].
\bibitem{pena}  H.~Garcilazo and M.~T.~Pena,
  Phys.\ Rev.\ C {\bf 72}, 014003 (2005)
  [arXiv:nucl-th/0406070] and references therein.
\bibitem{etanuctheo}
S.A.Rakityansky, S.A.Sofianos, V.B.Belyaev, and W.Sandhas
Phys. Lett. B {\bf 359} (1995) 33; S.~A.~Rakityansky, S.~A.~Sofianos, M.~Braun, V.~B.~Belyaev and W.~Sandhas,
  Phys.\ Rev.\ C {\bf 53} (1996) 2043;  A.~Fix and H.~Arenhovel,
  Phys.\ Rev.\ C {\bf 66} (2002) 024002.
\bibitem{winter}  P.~Winter {\it et al.},
  Phys.\ Lett.\ B {\bf 544}, 251 (2002)
  [Erratum-ibid.\ B {\bf 553}, 339 (2003)]
  [arXiv:nucl-ex/0210016].
\bibitem{arenh}
 A.~Fix and H.~Arenhoevel,
  Phys.\ Rev.\ C {\bf 69} (2004) 014001
  [arXiv:nucl-th/0310034].
\bibitem{veras}
  V.~Y.~Grishina  {\it et al.},
  Phys.\ Lett.\ B {\bf 475}, 9 (2000)
  [arXiv:nucl-th/9905049].
\bibitem{pawel}
 P.~Moskal,
  arXiv:hep-ph/0408162.
\bibitem{delhoff}  A.~Deloff,
  Phys.\ Rev.\ C {\bf 69} (2004) 035206
  [arXiv:nucl-th/0309059].
\bibitem{kanzo}
  K.~Nakayama, J.~Haidenbauer, C.~Hanhart and J.~Speth,
  Phys.\ Rev.\ C {\bf 68}, 045201 (2003)
  [arXiv:nucl-th/0302061].
\bibitem{report}
C. Hanhart, Phys. Rep. \textbf{397} (2004) 155 [arXiv:hep-ph/0311341].
\bibitem{thetapar}
 C.~Hanhart {\it et al.},
  Phys.\ Lett.\ B {\bf 590}, 39 (2004)
  [arXiv:hep-ph/0312236];  C.~Hanhart, J.~Haidenbauer, K.~Nakayama and U.~G.~Mei{\ss}ner,
  Phys.\ Lett.\ B {\bf 606}, 67 (2005)
  [arXiv:hep-ph/0407107].
\bibitem{ola}  R.~Frascaria {\it et al.},
  Phys.\ Rev.\ C {\bf 50} (1994) 537;  N.~Willis {\it et al.},
  Phys.\ Lett.\ B {\bf 406} (1997) 14
  [arXiv:nucl-ex/9703002];
 A.~Wronska {\it et al.},
  arXiv:nucl-ex/0510056 and these proceedings.
\bibitem{haider}Q. Haider and  L. C. Liu,
Phys. Rev. C {\bf 66} (2002) 045208;  Phys.\ Lett.\ B {\bf 172} (1986) 257.
\bibitem{phenoetanuc}  C.~Wilkin,
  Phys.\ Rev.\ C {\bf 47} (1993) 938
  [arXiv:nucl-th/9301006];
 S.~Wycech, A.~M.~Green and J.~A.~Niskanen,
  Phys.\ Rev.\ C {\bf 52} (1995) 544
  [arXiv:nucl-th/9502022].
\bibitem{green}
 A.~M.~Green and S.~Wycech,
  Phys.\ Rev.\ C {\bf 68} (2003) 061601
  [arXiv:nucl-th/0308057].
\bibitem{watson}
K. Watson, Phys. Rev. {\bf 88}, 1163 (1952);
A.B. Migdal, Sov. Phys. JETP {\bf 1}, 2 (1955). 
\bibitem{goldberger}
M. Goldberger and K. Watson, {\it Collision Theory} 
(Wiley, New York 1964).
\bibitem{achot2}
  A.~Gasparyan, J.~Haidenbauer and C.~Hanhart,
  Phys.\ Rev.\ C {\bf 72}, 034006 (2005)
  [arXiv:nucl-th/0506067].
\bibitem{saschas}
  A.~Sibirtsev, J.~Haidenbauer, C.~Hanhart and J.~A.~Niskanen,
  Eur.\ Phys.\ J.\ A {\bf 22} (2004) 495
  [arXiv:nucl-th/0310079].
\bibitem{mixingpaper}
 V.~Baru, J.~Haidenbauer, C.~Hanhart and J.~A.~Niskanen,
  Phys.\ Rev.\ C {\bf 68}, 035203 (2003)
  [arXiv:nucl-th/0303061].
\bibitem{recoils}
  V.~Baru, C.~Hanhart, A.~E.~Kudryavtsev and U.~G.~Mei{\ss}ner,
  Phys.\ Lett.\ B {\bf 589} (2004) 118
  [arXiv:nucl-th/0402027].
\bibitem{jounis}  J.~A.~Niskanen,
  arXiv:nucl-th/0508021.
\bibitem{sascha2}
 A.~Sibirtsev, J.~Haidenbauer, J.~A.~Niskanen and U.~G.~Mei{\ss}ner,
  Phys.\ Rev.\ C {\bf 70} (2004) 047001
  [arXiv:nucl-th/0407073].
\bibitem{askalbrecht}A. Gillitzer et al., COSY proposal 102 (2001) and
  contribution to these proceedings.
\bibitem{magiera}
  A.~Magiera and H.~Machner,
  Nucl.\ Phys.\ A {\bf 674} (2000) 515.
\end{thebibliography}
\end{document}